\documentclass[letterpaper, 10 pt, conference]{ieeeconf} 
\IEEEoverridecommandlockouts                    
\let\labelindent\relax
\usepackage{enumitem}

\usepackage{mathtools,amsthm}

\usepackage{graphics} 
\usepackage{epsfig}
\usepackage{times} 
\usepackage{amsmath,amssymb,amsfonts}
\usepackage{mathrsfs}
\usepackage{algorithmic}
\usepackage{graphicx}
\usepackage{textcomp}
\usepackage{varioref}
\usepackage{import}
\usepackage{framed,xcolor}
\usepackage{color}
\usepackage{comment}
\usepackage{multirow}
\usepackage{epstopdf}   
\usepackage{psfrag}
\usepackage{breqn}
\usepackage{float}
\usepackage{mathtools}
\usepackage{booktabs}
\usepackage{soul}
\usepackage{amsbsy}
\usepackage[bottom]{footmisc}
\usepackage{lineno}
\usepackage{cleveref}
\usepackage{microtype}
\usepackage{comment}

\definecolor{arash}{rgb}{0.8,0.8,1}
\definecolor{seb}{rgb}{0.8,1,0.8}
\definecolor{seb2}{rgb}{0.5,.5,1}
\definecolor{arash2}{rgb}{0,.5,0}
\definecolor{wenqi}{rgb}{1,.75,0.79}
\definecolor{wenqi2}{rgb}{1,.75,0.79}

\graphicspath{{Figures/}}
\newcommand{\vect}[1]{\ensuremath{\boldsymbol{\mathrm{#1}}}}
\usepackage{graphicx}
\usepackage{tabularx,booktabs}
\usepackage{color}
\usepackage{multicol}
\usepackage{amsmath,amssymb}
\usepackage{amsfonts}
\usepackage{textcomp}
\usepackage{psfrag}
\usepackage{multimedia}
\usepackage{fancybox}
\usepackage{comment}
\usepackage{soul}
\makeatletter
\newcommand{\biggg}{\bBigg@{1.6}}  

\makeatother

\definecolor{seb}{rgb}{0.8,1,0.8}
\definecolor{arash}{rgb}{0.8,0.8,1}

\newcounter{lastnote}

\DeclareMathOperator*{\argminA}{arg\,min} % Jan Hlavacek

\title{\LARGE \bf
Reinforcement Learning based on Scenario-tree MPC for ASVs
}
\author{Arash Bahari Kordabad, Hossein Nejatbakhsh Esfahani, Anastasios M. Lekkas, Sébastien Gros
\thanks{The authors are with Department of Engineering Cybernetics, Norwegian University of Science and Technology (NTNU), Trondheim, Norway. E-mail:{\tt\small \{Arash.b.kordabad, hossein.n.esfahani, anastasios.lekkas, sebastien.gros\}@ntnu.no}}
 }
\begin{document}
\maketitle
\thispagestyle{empty}
\pagestyle{empty}
%%%%%%%%%%%%%%%%%%%%%%%%%%%%%%%%%%%%%%%%%%%%%%%%%%%%%%%%%%%%%%%%%%%%%%%%%%%%%%%%
\begin{abstract}
In this paper, we present the use of Reinforcement Learning (RL) based on Robust Model Predictive Control (RMPC) for the control of an Autonomous Surface Vehicle (ASV). The RL-MPC strategy is utilized for obstacle avoidance and target (set-point) tracking. A scenario-tree robust MPC is used to handle potential failures of the ship thrusters. Besides, the wind and ocean current are considered as unknown stochastic disturbances in the real system, which are handled via constraints tightening. The tightening and other cost parameters are adjusted by RL, using a Q-learning technique. An economic cost is considered, minimizing the time and energy required to achieve the ship missions.
The method is illustrated in simulation on a nonlinear 3-DOF model of a scaled version of the Cybership \MakeUppercase{\romannumeral 2}.
\end{abstract}
\section{INTRODUCTION}
Autonomous Surface Vehicles (ASVs) have been extensively investigated recently in industry and research~\cite{ESFAHANI2019106526,martinsen2020optimization,bitar2020two}. However, designing control systems that can tackle obstacle avoidance and tracking control, with severe external time-varying disturbances due to the wind, wave, and ocean currents, is one of the most challenging research topics for ASVs in maritime engineering~\cite{WOO2019155,martinsen2020reinforcement}. In the control literature, the motion control scenarios of such vehicles are divided into target tracking, path following, path tracking, and path maneuvering~\cite{breivik2010topics}.  This paper focuses on target (set-point) tracking motion control in the presence of static elliptic-shape obstacles and mission-varying wind and ocean current. In set-point tracking, only a terminal point is given, which ought to be reached at minimum cost.
\par Reinforcement Learning (RL) is a powerful tool for tackling Markov Decision Processes (MDP) without prior knowledge of the process to be controlled~\cite{bertsekas2019reinforcement}. Indeed, RL attaches a reward function to each state-action pair and tries to find a policy to optimize the discounted infinite rewards labelled performance~\cite{sutton2018reinforcement}. Dynamic Programming (DP) methods can be used to solve MDPs. However, DP requires a knowledge of the MDP dynamics, and its computational complexity is unrealistic in practice for systems having more than a few states and inputs. Instead, most investigations in RL have focused on achieving approximate solutions, while not requiring a model of the dynamics. Fuzzy Neural Networks and Deep Neural Networks (DNNs) are a common choice to approximate the optimal policy~\cite{ELearning}.
However, analysing formally the closed-loop behavior of a learned policy based on a DNN, such as stability and constraints satisfaction is challenging. Moreover, providing meaningful initial weights for the DNN can be very difficult.
 For instance, in~\cite{zhang2020modelreference} the baseline control is employed to ensure stability and tracking performance of ASV, while DNN-based RL is added to handle uncertainties and collision avoidance.
\par Model Predictive Control (MPC) is a well-known model-based control method that employs a model of the system dynamics to build an input sequence over a given finite horizon such that the resulting predicted state trajectory minimizes a given cost function while respecting the constraints imposed on the system~\cite{MPCbook}. The first input is applied to the real system, and the problem is solved at each time instant based on the latest state of the system. The advantage of MPC is its ability to explicitly support state and input constraints, while producing a nearly optimal policy~\cite{wang2019parallel}. However, model uncertainties can severely impact the performance of the MPC policy.
\par In Robust Model Predictive Control (RMPC), Scenario- tree MPC is a useful approach to handle nonlinear systems with finite and discrete uncertainties. Scenario-based MPC approach for ship collision avoidance is presented in~\cite{johansen2016ship}. Tube-based MPC is another technique for RMPC mostly used when the MPC model and constraints are linear and the uncertainties can be contained in a polytope~\cite{MPCbook}.
\par Data-driven adaptation of the RMPC model, e.g. using system identification, to better fit the real system is a fairly obvious strategy to tackle the issues concerning inaccurate model and unknown disturbance. However, if the model cannot capture the real system dynamics, adapting the model from data does not necessarily improve the performance of the MPC policy.
Instead, we propose to use RL to online tune the RMPC formulation using the data obtained from the real system~\cite{gros2019data}. Unlike DNN, MPC as a function approximator for RL, can explicitly handle constraints satisfaction, stability, and safety~\cite{Zanon2019b,gros2020reinforcement,Arash2021MPC,Arash2021ACC}. 
\par In this paper, we use a scenario-tree MPC to manage potential thruster failures. Constraint tightening is used to avoid the obstacles in the presence of stochastic wind and ocean current. We consider a trade-off between time and energy to reach a neighborhood of the target as a baseline cost of RL. This cost is penalized linearly when approaching the obstacles. RL will adjust the tightening parameter and other RMPC parameters to find an optimal policy during some missions. 
\par The paper is structured as follows. Section \ref{sec:model} presents the 3-DOF nonlinear ship's dynamics and its thruster configuration. Section \ref{sec:RMPCRL}  formulates the scenario-tree MPC and RL, and details an RMPC parameterized scheme as a function approximator of Q-learning. Section \ref{sec:sim} describes the simulation details and illustrate the  results. The target point tracking with back-off constraint in obstacle will be considered, and Q-learning tunes the parameters.
\section{Vessel Model}\label{sec:model}
\begin{figure}[ht!]
		\centering
	{\def\svgwidth{0.45\textwidth}
			\small
			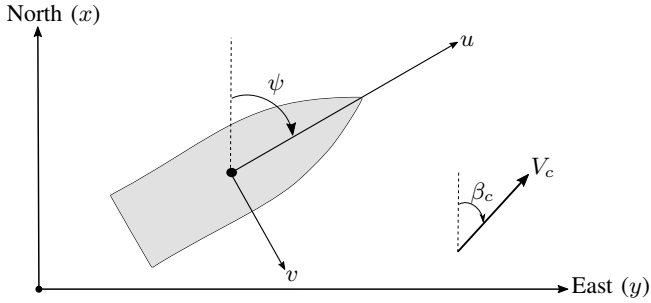
			\normalsize}
	\caption{ The 3-DOF Ship model in North-East-Down (NED) frame with surge $u$, sway $v$ and heading angle $\psi$, and ocean current vector $v_c$}
	\label{figdyn}
\end{figure}

The 3-DOF nonlinear dynamics of the Cybership \MakeUppercase{\romannumeral 2} can be represented by a pose vector  $\vect{\eta}=[x,y,\psi]^{\top}\in\mathbb{R}^3$ in the Earth-fixed frame, where $x$ is the North position, $y$ is the East position, $\psi$ is the heading angle. The velocity vector $\vect{\nu}=[u,v,r]^{\top}\in\mathbb{R}^3$ includes the surge $u$ and sway $v$ velocities, and yaw rate $r$ decomposed in the body-fixed frame (see Fig.\ref{figdyn}). The model dynamics can be written as follows~\cite{Cybership}:
\begin{subequations}\label{con-dyn}
\begin{align}
   &\qquad\qquad\qquad\qquad\qquad\dot{\vect{\eta}}={J}(\psi)\vect{\nu} \\
    & M_{RB}{\dot{\vect\nu}}+M_{A}{\dot{\vect\nu}_r}+C_{RB}({\vect\nu}){\vect\nu}+ C_{A}({\vect\nu_r}){\vect\nu_r}+\nonumber\\&\qquad\qquad\qquad\qquad\qquad \qquad\quad
     D({\vect\nu_r}){\vect\nu_r}=\vect\tau + \vect\tau_w
\end{align}
\end{subequations}
where ${\vect\nu}_{r}={\vect\nu}-{\vect\nu}_c=[u_r,v_r,r]^\top$ is the ship velocity relative to the ocean current, and ${\vect\nu}_c={J}(\psi)^\top \vect v_c$ where $
\vect v_c=[V_{c}\cos{\beta_c},V_{c}\sin{\beta_c},0]^{\top}$ are the ocean current in the body-fixed and Earth-fixed frames, respectively, and where $V_{c}$ is the current velocity and $\beta_c$ is its angle in the Earth-fixed frame. The rotation matrix ${J}(\psi)$ is given by:
\begin{gather}
    {J}(\psi)=
    \begin{bmatrix}
    \cos(\psi)& -\sin(\psi) &0 \\ 
    \sin(\psi)& \cos(\psi) &0\\ 
    0& 0 & 1
    \end{bmatrix}
\end{gather}
The rigid-body inertia matrix $M_{RB}$ and added mass $M_A$ are given by:
\begin{gather}
    M_{RB}=
    \begin{bmatrix}
    m & 0  & 0 \\ 
    0& m  &mx_g\\ 
    0& mx_g &I_z
    \end{bmatrix}
    ,\,\, M_A=\begin{bmatrix}
    -X_{\dot{u}} & 0 & 0\\ 
    0& -Y_{\dot{v}}  & -Y_{\dot{r}}\\ 
    0& -N_{\dot{v}} & -N_{\dot{r}}
    \end{bmatrix}
\end{gather}
where $m$ is the mass of the ship, $I_z$ is the moment
of inertia about the body $z_b$-axis (yaw axis) and $x_g$ is the distance between the centre of gravity and the body $x_b$-axis.
Furthermore, the rigid-body and hydrodynamic of the Centripetal and Coriolis acceleration matrices read as:
\begin{subequations}
\begin{gather}
    \mathbf{C}_{RB}(\boldsymbol{\nu})=\begin{bmatrix}
        0 & 0 & -m(x_{g}r+v) \\ 
        0& 0  & mu\\ 
        m(x_{g}r+v)& -mu &0
    \end{bmatrix}\\
    \mathbf{C}_{A}(\boldsymbol{\nu}_r)=
    \begin{bmatrix}
        0 & 0& c_{13} \\ 
        0& 0&  c_{23}\\ 
        -c_{13}& -c_{23}& 0
    \end{bmatrix}
\end{gather}
\end{subequations}
where $c_{13}=Y_{\dot{v}}v_r+0.5(N_{\dot{v}}+Y_{\dot{r}})r$, $c_{23}=-X_{\dot{u}}u_r$, and $X_{\dot{u}}$, $Y_{\dot{v}}$, $Y_{\dot{r}}$, $N_{\dot{v}}$ and $N_{\dot{r}}$ are constant model parameters~\cite{fossen2011handbook}.
Moreover, the damping matrix is:
\begin{subequations}\label{first:main}
	\begin{gather}
    	\mathbf{D}(\boldsymbol{\nu}_r)=-
    	\begin{bmatrix}
    	d_{11} & 0 & 0 \\ 
    	0& d_{22} &d_{23}\\ 
    	& d_{32} &d_{33}
    	\end{bmatrix}\tag{\ref{first:main}}
    \end{gather}
	where
	\begin{align}
	d_{11}=& X_u+X_{|u|u}|u_r|+X_{uuu}u_r^2\\
	d_{22}=& Y_v+Y_{|v|v}|v_r|+Y_{|r|v}|r|\\
	d_{23}=& Y_r+Y_{|v|r}|v_r|+Y_{|r|r}|r|\\
	d_{32}=& N_v+N_{|v|v}|v_r|+N_{|r|v}|r|\\
	d_{33}=& N_r+N_{|v|r}|v_r|+N_{|r|r}|r|
	\end{align}
\end{subequations}
where $X_{(.)}$, $Y_{(.)}$, and $N_{(.)}$ are the hydrodynamic
coefficients~\cite{fossen2011handbook}.
The model parameters are taken from~\cite{Cybership}.
Finally, $\vect\tau=[X,Y,N]^{\top}$ is the external control forces $X, Y$ and moment $N$ vector and $\vect\tau_w$ is the wind effects disturbance.
\subsection{Thruster Allocation}
\begin{figure}[ht!]
	\centering
	\def\svgwidth{0.48\textwidth}
	\small
	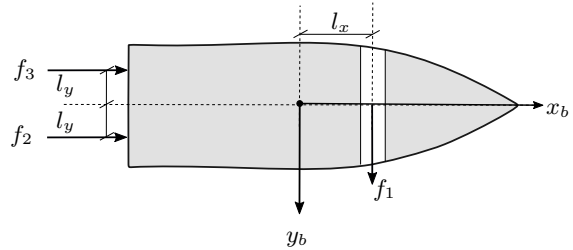
	\normalsize
	\caption{Schematic drawing showing the thrusters configuration in the body-fixed frame $\left \{ b \right \}$ }
	\label{fig:ship}
\end{figure}

We consider one tunnel thruster (transverse) $f_1$ and two main propeller thrusters (longitudinal) $f_2,f_3$ as the thrust configuration (see Fig. \ref{fig:ship}). Then
\begin{gather}\label{eq:tau}
    \vect \tau=
    \begin{bmatrix}
    0 & 1 & 1 \\ 
    1& 0 & 0\\
    l_x& -l_y &l_y
    \end{bmatrix}\vect a
\end{gather}
where $\vect a=[f_1,f_2,f_3]^\top$ is the actuator forces vector subject to the bounds:
\begin{gather}
\vect a_{\mathrm{min}} \leq \vect a\leq \vect a_{\mathrm{max}}  \label{aleq0}
\end{gather}
\subsection{Obstacle Avoidance}
For simplicity, we consider obstacles of elliptic shape. Hence, the condition for obstacles avoidance can be seen as the following inequality:
\begin{gather}
\left(\left(x-o_{x,j}\right)/\left(r_{x,j}+r_o\right)\right)^2+\left(\left(y-o_{y,j}\right)/\left(r_{y,j}+r_o\right)\right)^2\geq 1 \label{obs}
\end{gather}
where $(o_{x,j},o_{y,j})$ and $(r_{x,j},r_{y,j})$  are the center and radii of the $j^{\mathrm{th}}$ ellipse ($j=1,\ldots,N_o$), respectively, $r_o$ is radius of the vessel and $N_o$ is number of obstacles.
\section{RMPC-based Reinforcement learning} \label{sec:RMPCRL}
In this section, we formulate the scenario-tree MPC scheme and detail how it can be treated via Q-learning.
\subsection{Robust Model Predictive Control}
\begin{figure}[ht!]
	\centering
	\includegraphics[width=0.45\textwidth]{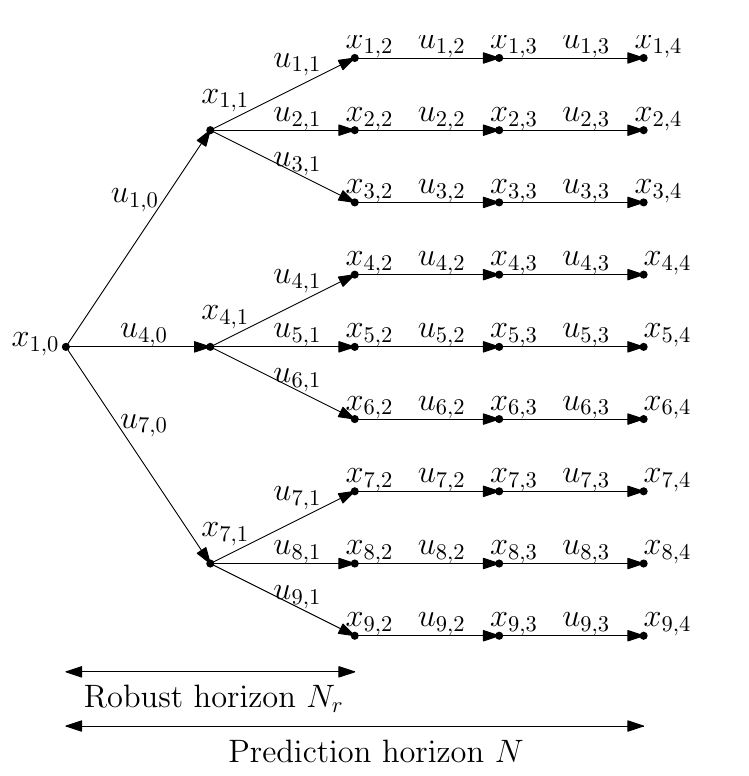}
	\caption{The evolution of the system represented as a scenario tree~\cite{klintberg2016improved}}
	\label{scenario tree}
\end{figure}

Scenario-tree MPC is a robust MPC technique that can treat finite and discrete uncertainties in the system~\cite{klintberg2016improved}. Fig. \ref{scenario tree} shows the evolution of the system described by a scenario tree, where $\vect x_{k,i}$ and $\vect u_{k,i}$ are the state and input of scenario $k$ at time $i$, given by:
\begin{align}
   \vect x_{k,i+1}=\vect f_{k,i}\left(\vect x_{k,i},\vect u_{k,i}\right)
\end{align}
where $\vect f_{k,i}$ is the $k^{\mathrm{th}}$ (time-varying)  model. In this paper, the scenario tree will be used to model the thruster failures in the system, hence each model $\vect f_{k,i}$ corresponds to a specific failure $k$ at a specific time $i$.
Since the number of scenarios grows exponentially with the length of the MPC horizon, it is common to fix the uncertain parameters after a certain period of time called Robust horizon $N_r<N$, where $N$ is the MPC prediction horizon. Then the number of scenarios is $M=m_d^{N_r}$, where $m_d$ is the number of realization (branches) at each time stage.
We assumed separate state and control variables for each scenario to enable parallel computations. However, because the uncertainty cannot be anticipated, control action must depend on only the historical realizations of the uncertainty. Then, $\vect u_{k,j}=\vect u_{l,j}, \forall j=0,...,i$ if the uncertainty realization for scenario $k$ and $l$ are identical up to and including the time stage $i$. This restriction is commonly denoted as non-anticipativity constraint. In Fig. \ref{scenario tree}, $N=4$, $m_d=3$, $N_r=2$ and then, $M=m_d^{N_r}=9$. Also,  $\vect u_{1,0}=\vect u_{2,0}=\vect u_{3,0},\vect u_{4,0}=\vect u_{5,0}=\vect u_{6,0},\vect u_{7,0}=\vect u_{8,0}=\vect u_{9,0}$ are the non-anticipativity constraints.
\subsection{Reinforcement Learning}
Reinforcement Learning considers that the real system is described by a Markov Decision Process (MDP) with state transitions having the
underlying conditional probability density $
    \mathbb{P}\left[\vect s_+|\vect s,\vect a\right]$,
where $\vect s,\vect a$ is the current state-input pair and $\vect s_+$ is the
subsequent state. The control literature typically uses the notation $\vect s_{+}=f^{\mathrm{real}}({\vect s},{\vect a},{\vect \zeta})$, where $\vect \zeta$ is a random disturbance and $\vect f^{\mathrm{real}}$ is discretized real system dynamics \eqref{con-dyn} and $\vect s=[\vect\eta^\top,\vect\nu^\top]^\top$.
We will label $L(\vect s,\vect a)$ as the baseline stage cost associated to the MDP at each transition. The optimum action-value function $Q_{\star}$, optimum value function $V_{\star}$ and optimum policy $\vect \pi_{\star}$ associated to the MDP are defined by the Bellman equations:
\begin{subequations}
\label{eq:QlearningRule}
	\begin{align}
	V_{\star}(\vect s) &= \min_{\vect a} Q_{\star}(\vect s,\vect a),\\
	Q_{\star}(\vect s,\vect a) &= L(\vect s,\vect a) + \gamma \mathbb{E}[V_{\star}(\vect s_{+})|\vect s,\vect a],\\
	\vect \pi_{\star}(\vect s)&= \argminA_{\vect a}Q_{\star}(s,a)
	\end{align}
\end{subequations}
where $\gamma \in (0,1]$ is the MDP discount factor. 
\par Q-learning is a classical model-free RL algorithm that tries to capture the action value function $Q_{\vect \theta}\approx Q_\star$ via tuning the parameters vector $\vect \theta \in \mathbb{R}^n$. The approximation of the value function $V_{\vect \theta}$ and parametric optimal policy $\vect \pi_{\vect \theta}$ can then be extracted from the Bellman equations. Q-learning uses the following update rule for the parameters $\vect \theta$ at state $\vect s_k$~\cite{RLAlgorithms}:
\begin{subequations}\label{eq:Qlearning}
	\begin{gather}
	\delta_{k}=L(\vect s_{k},\vect a_{k})+\gamma V_{\vect \theta}(\vect s_{k+1})-Q_{\vect \theta}(\vect s_{k},\vect a_{k})\\
	\vect \theta\leftarrow \vect \theta+\alpha\delta_{k}\nabla_{\vect \theta}Q_{\vect \theta}(\vect s_{k},\vect a_{k}) \label{Qlearning}
	\end{gather}
\end{subequations}
where the scalar $\alpha>0$ is the learning step-size, $\delta_k$ is labelled the Temporal-Difference (TD) error and
the input $\vect a_k$ is selected according to the corresponding parametric policy $\vect \pi_{\vect \theta}(\vect s_k)$ with possible addition of small random exploration.
\par Using RMPC as a way of supporting the approximations $V_{\vect \theta}$ and $Q_{\vect \theta}$ has been proposed and justified in~\cite{gros2019data}. Hereafter, we detail how this can be done for the specific choice of RMPC proposed here.
\subsection{RMPC as a function approximator for RL}
We propose to use the action-value function approximate $Q_{\vect \theta}\approx Q_\star$  obtained from the following RMPC scheme parameterized by $\vect \theta$~\cite{gros2019data}:
\begin{subequations}\label{MPC}
\begin{align}
	Q_{\vect \theta}(\vect s,\vect a)=\min_{\vect {x},\vect {u},\vect {\sigma}} &\,\,\sum_{k=1}^{M} \Bigg( \gamma^{N}V_{k}^{f}\left(\vect x_{k,N},\vect \theta\right)+\vect \omega_{f}^\top\vect \sigma_{k,N}+  \Bigg.\label{cost}\nonumber\\ \Bigg. \sum_{i=0}^{N-1}&\left(\gamma^{i}l_{k}\left(\vect x_{k,i},\vect u_{k,i},\vect \theta\right)+\vect \omega^\top\vect \sigma_{k,i}\right)\Bigg)\\
	\mathrm{s.t.}& \,\, \forall i=0,...,N-1,\forall k=1,...,M:\nonumber\\ 
	& \,\, \vect x_{k,i+1} = \vect f_{k,i}(\vect x_{k,i},\vect u_{k,i},\vect \theta) \label{con-e}\\
     & \,\, \vect h_{\vect \theta}(\vect x_{k,i},\vect u_{k,i})+\vect B_{k,i}(\vect\theta)\leq \sigma_{k,i} \label{con-si1i}\\
     & \,\, \vect h_{\vect \theta}^{f}(\vect x_{k,N})+\vect B^f_{k,N}(\vect \theta)\leq \sigma_{k,N} \label{con-si2i}\\
	& \,\, \vect g(\vect u_{k,i})\leq 0 \label{con-ii}\\ 
	& \,\, \vect u_{k,i}=\vect u_{l,i} \text{ if } \vect x_{k,j}=\vect x_{l,j},\label{non-ant} \nonumber \\
	& \,\, \forall k,l \in \left \{ 1,..,M \right \}, \forall j \in \left \{ 1,...,i \right \}  \\
	& \,\, \vect x_{k,0}=\vect s\label{state-cons}\\ 
	& \,\, \vect u_{k,0}=\vect a \label{action-cons}\\
		& \,\, \vect \sigma_{k,i}\geq 0 , \quad \vect \sigma_{k,N}\geq 0 \label{con-sig} 
	\end{align}
\end{subequations}
where $\vect {x}=\left\{\vect x_{1,0},\ldots ,\vect x_{M,N}\right\}$, $\vect {u}=\left\{\vect u_{1,0},\ldots,\vect u_{M,N-1}\right\}$ and $\vect {\sigma}=\left\{\vect \sigma_{1,0},\ldots ,\vect \sigma_{M,N}\right\}$  are the primal decision variables,
$M$ is the number of scenarios, $N$ is the prediction horizon, $\vect f_{\{1,\ldots ,M\},i}$ are $M$ different (possibly) time-varying models supporting the discrete uncertainties,  $l_{1,\ldots ,M}$ and $V_{1,\ldots ,M}^{f}$ are the stage and terminal costs for the different scenarios, respectively. The constraint tightening is performed in \eqref{con-si1i} and \eqref{con-si2i}, where $\vect B_{k,i}(\vect\theta)\geq 0$ and $\vect B^f_{k,N}(\vect\theta)\geq 0$ are the (possibly) time-varying tightening parameters. Variables $\vect \sigma_{k,i}$ and $\vect \sigma_{k,N}$ are slacks for the relaxation of the mixed state-input constraints, using the positive weights vectors $\vect \omega$ and $\vect \omega_f$, respectively. The relaxation prevents the infeasibility of the tightened constraints of the RMPC in the presence of disturbances and mismatching models $\vect f_{k,i}$ to the real system $\vect f^{\mathrm{real}}$. Constraint \eqref{con-ii} represents the input inequality constraints which is defined in \eqref{aleq0} for the ASV. Constraint \eqref{non-ant} is the non-anticipativity constraint~\cite{lucia2014handling}
\par In \eqref{MPC}, $\vect \theta$ is the parameters vector that can be modified by RL to shape the action-value function. Under some mild assumptions~{(see~\cite{gros2019data} for the technical details)}, if the parametrization is rich enough, the MPC scheme is able to capture the true optimal action-value function $Q_\star$, value function $V_\star$ and policy $
\vect \pi_\star$ jointly, even if the RMPC models $\vect f_{k,i}$ do not capture the real system dynamics \eqref{con-dyn}. 
\par
One can verify that the parameterized value function $V_{\vect \theta}$ that satisfies the Bellman equations can be obtained by solving \eqref{MPC} without constraint \eqref{action-cons}. Moreover, the parameterized deterministic policy $\vect \pi_{\vect \theta}$ reads as follows: 
 \begin{gather}
    \vect \pi_{\vect \theta}(\vect s)=\vect u_{k,0}^{\star}(\vect s,\vect \theta)
\end{gather}
where $\vect u_{k,0}^{\star}(\vect s,\vect \theta)$ is the first element of $\vect {u}^\star$, solution of the RMPC scheme \eqref{MPC} when constraint \eqref{action-cons} is removed.
Therefore, the value function $V_{\vect \theta}(\vect s)$ can be acquired together with the policy $  \vect \pi_{\vect \theta}(\vect s)$ by solving a classic MPC scheme, while the action value function results from solving the same MPC scheme with its first input constrained to a specific value $\vect a$.
\par The sensitivity $\nabla_{\vect \theta}Q_{\vect \theta}(\vect s,\vect a)$ required in \eqref{Qlearning} is given by~\cite{gros2019data}:
\begin{gather}
    \nabla_{\vect \theta}Q_{\vect \theta}(\vect s,\vect a)=\nabla_{\vect \theta}\mathcal{L}_{\vect \theta}(\vect s,\vect a,\vect y^{\star})
\end{gather}
where $\mathcal{L}$ is the Lagrange function associated to the scenario-tree RMPC \eqref{MPC}, i.e.:
\begin{align}
\mathcal{L}_{\vect \theta}(\vect s,\vect a,\vect {y}) = \Phi_{\vect \theta} + \vect \lambda^\top \vect G_{\vect \theta}  + \vect \mu^\top \vect H_{\vect \theta} 
\end{align}
where $\Phi_{\vect \theta}$ is the cost \eqref{cost}, $\vect G_{\vect \theta}$ gathers the equality constraints \eqref{con-e}, \eqref{non-ant}, \eqref{state-cons}, \eqref{action-cons}, $\vect H_{\vect \theta}$ collects the inequalities \eqref{con-si1i}, \eqref{con-si2i}, \eqref{con-ii}, \eqref{con-sig}, and $\vect \lambda,\vect \mu$ are the associated dual variables. Argument $\vect {y}$ reads as $\vect {y} = \left\{\vect {x},\vect {u},\vect {\sigma},\vect \lambda,\vect \mu\right\}$ and  $\vect  {y}^{\star}$ is the solution to \eqref{MPC}.
\section{Simulation}\label{sec:sim}
In this section, we consider a target tracking problem in the presence of static obstacles modelled as ellipsoids, random wind and ocean currents, and discrete uncertainties in the dynamics. The objective is to reach the terminal (target) point while achieving an optimal trade-off between time and energy.
\par
We consider the nominal system as the first scenario ($k=1$) and the failure of thrusters $f_2$ or $f_3$ ($k=2,3$) as the discrete uncertainties in the system. As a result, by considering $N_r=1$, this formulation has $M=m_{d}=3$ scenarios and realization at each time instance.
\par We consider a stage cost that minimizes both the energy and time.  Also, the stage cost is in the form:
\begin{gather}
    L(\vect s,\vect a)=\underbrace{|X.u|+|Y.v|+|N.r|}_{\text{power}}+\underbrace{T}_{\text{time }}+\underbrace{\vect {c}^\top\max(0,\vect h_{\vect \theta}+\vect d)}_{\text{obstacles penalty}}
\end{gather}
where $T$ is a constant introducing a penalty on the time to reach the target. The term $\vect {c}^\top \max(0,\vect h_{\vect \theta}+\vect d)$ penalizes violations of the relaxed inequality constraints $\vect h_{\vect \theta} + \vect d\leq 0$ with a weight vector $\vect {c}$. The parameter $\vect d$ can be interpreted as the dangerous distance from the obstacles. Indeed, when $ 0 < \vect h_{\vect \theta}+\vect d$, RL tries to increase the distance by adjusting the MPC tightening parameters.  Since the task is episodic here, we can use an undiscounted cost in RL i.e. $\gamma =1$.
\par The obstacles constraints tightening is parametrized as follow:
\begin{gather}
  \vect B_{k,i}(\vect \theta)=\vect B^f_{k,N}(\vect \theta)=\vect \theta_k^h
\end{gather}
where $\vect \theta_k^h=\vect \theta_{k,\{1,\ldots,N_o\}}^h$ is horizon-invariant parameter and we use $N_o=2$ obstacles. 
\par The stochastic ocean current is represented as $\vect \zeta=\left \{ V_c,\beta_c \right \}$. We generate a random current map for each mission independently, using the gradient of Gaussian Radial Basis Functions set, as follows:
\begin{align}
    \vect {\upsilon}_c= \frac{\partial}{\partial \vect {p}} \sum_{l=1}^{N_c} q_{l} \exp{\left(-\frac{\|\vect {p}-\vect {b}_l\|^2}{2\rho^2_l}\right)}
\end{align}
where $\vect {p}=[x,y]^\top$ is the position vector, $\{ q_l, \vect {b}_l, \rho_l \}$ are random values and $N_c$ is the number of Gaussian functions which we take $N_c=2$ here. Then $V_c$ and  $\beta_c$ are obtained as magnitude and angle of the vector $\vect {\upsilon}_c$.
\par We consider $N=20$ the prediction horizon. A sampling time of $\text{dt}=0.5$s was chosen for the discretization of the system dynamics \eqref{con-dyn}, and the actuators bounds as $\vect a_{\mathrm{max}}=[2,8,8]^\top\mathrm{N}$ and $\vect a_{\mathrm{min}}=[-2,0,0]^\top\mathrm{N}$ in \eqref{aleq0}. In addition, the stage and terminal costs of the RMPC scheme can be represented as the following weighted vector norm:
\begin{subequations}
\begin{align}
    l_k(\vect x_{k,i},\vect u_{k,i},\vect \theta)&=\left\|\left[\left(\vect x_{k,i}-\vect X_{ref}\right)^\top, \vect u^\top_{k,i}\right]^\top\right\|_{ \Theta_k^l}\\
    V_k^f(\vect x_{k,i},\vect \theta)&=\left\|\vect x_{k,i}-\vect X_{ref}\right\|_{ \Theta_k^V}    
\end{align}
\end{subequations}
where $\vect X_{ref}$ is the reference state in the target-tracking and parameters $ \Theta_k^l$ and $ \Theta_k^V$ are the weights of the vector norm. They can be tune by RL as well. The RL parameters read as:
\begin{gather} 
    \vect \theta= \left\{\vect \theta^h,  \Theta_1^l,\ldots, \Theta_M^l,\Theta_1^V,\ldots,\Theta_M^V \right\}
\end{gather}

\begin{figure}[ht!]
	\centering
	\def\svgwidth{0.3\textwidth}
	\includegraphics[width=0.48\textwidth]{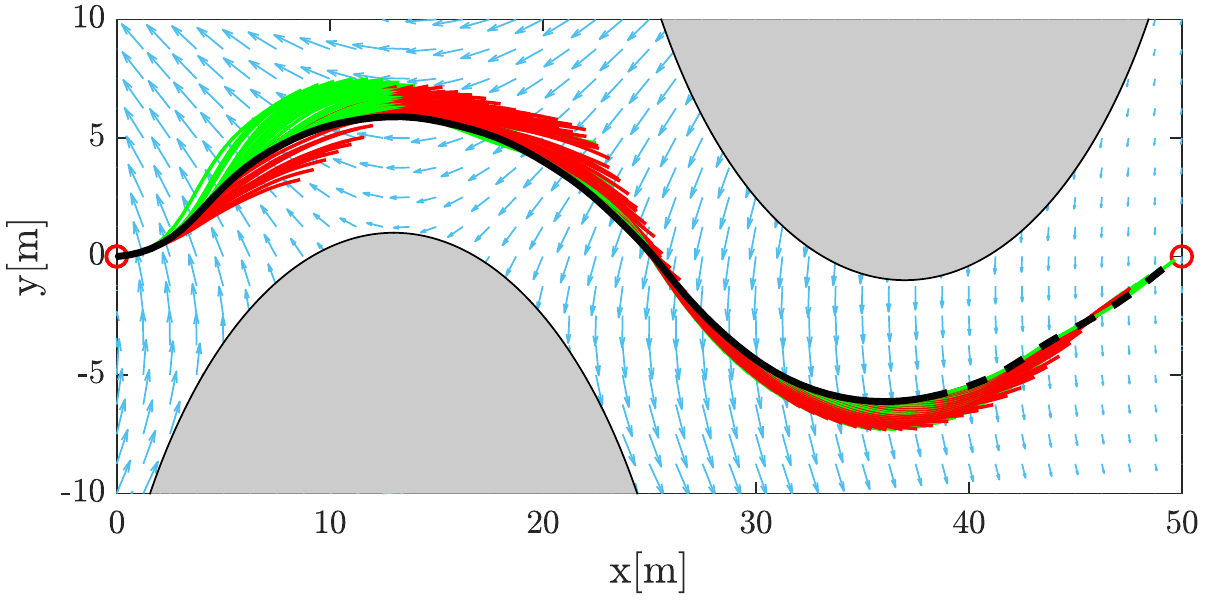}
	\caption{The path of the ship (black) in the first mission and random current, fail prediction ($k=2,3$): red and nominal prediction ($k=1$): green. RL updating is stopped in the dashed-line (the MPC prediction at the end of missions).}
	\label{result1}
\end{figure}

Fig. \ref{result1} shows the path for the first simulated mission. The corresponding random wind and ocean current map is shown as well. The failure scenario prediction and nominal scenario are specified by red and green, respectively. The learning process continues until RMPC predicts the target point as the terminal state for the first time. Once the target point is within the RMPC horizon, a different control scheme ought to be used.

\begin{figure}[ht!]
	\centering
	\def\svgwidth{0.3\textwidth}
	\includegraphics[width=0.48\textwidth]{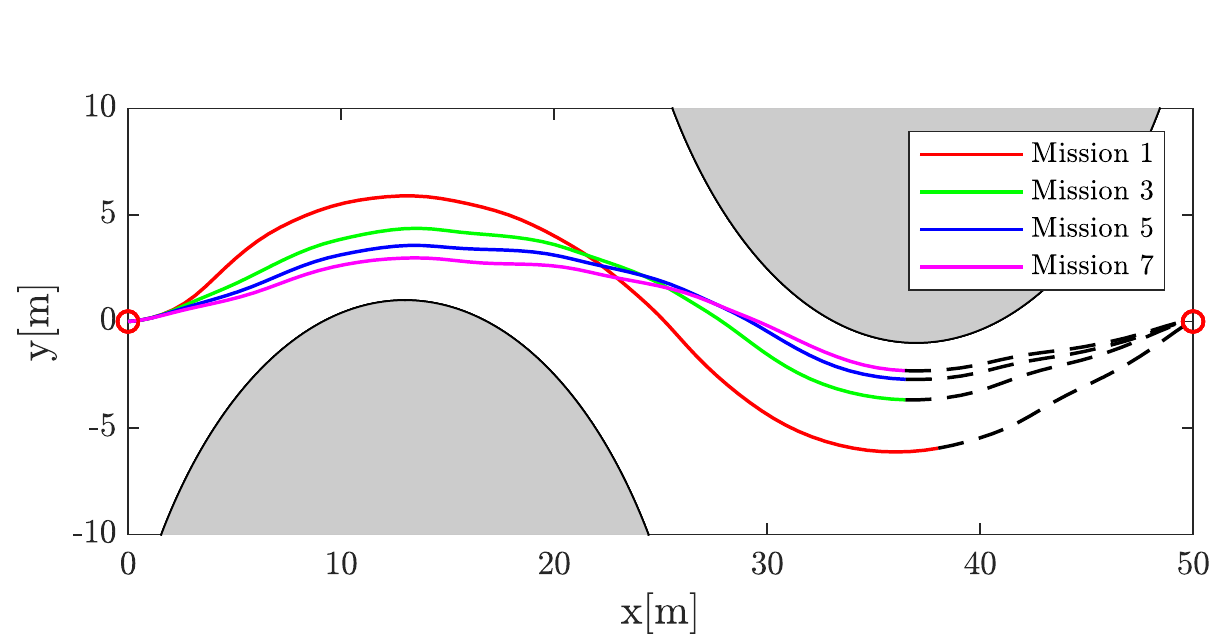}
	    \caption{The path of the ship over missions.}

	\label{result2}
\end{figure}

Fig. \ref{result2} illustrates the paths over missions for the nominal system. We simulated seven missions and for the sake of brevity four missions were selected for illustration. It can be seen that the paths are nearing to obstacles during the missions until the  RL penalty is  activated and find the optimal distance to handle disturbance.

\begin{figure}[ht!]
	\centering
	\def\svgwidth{0.3\textwidth}
	\includegraphics[width=0.48\textwidth]{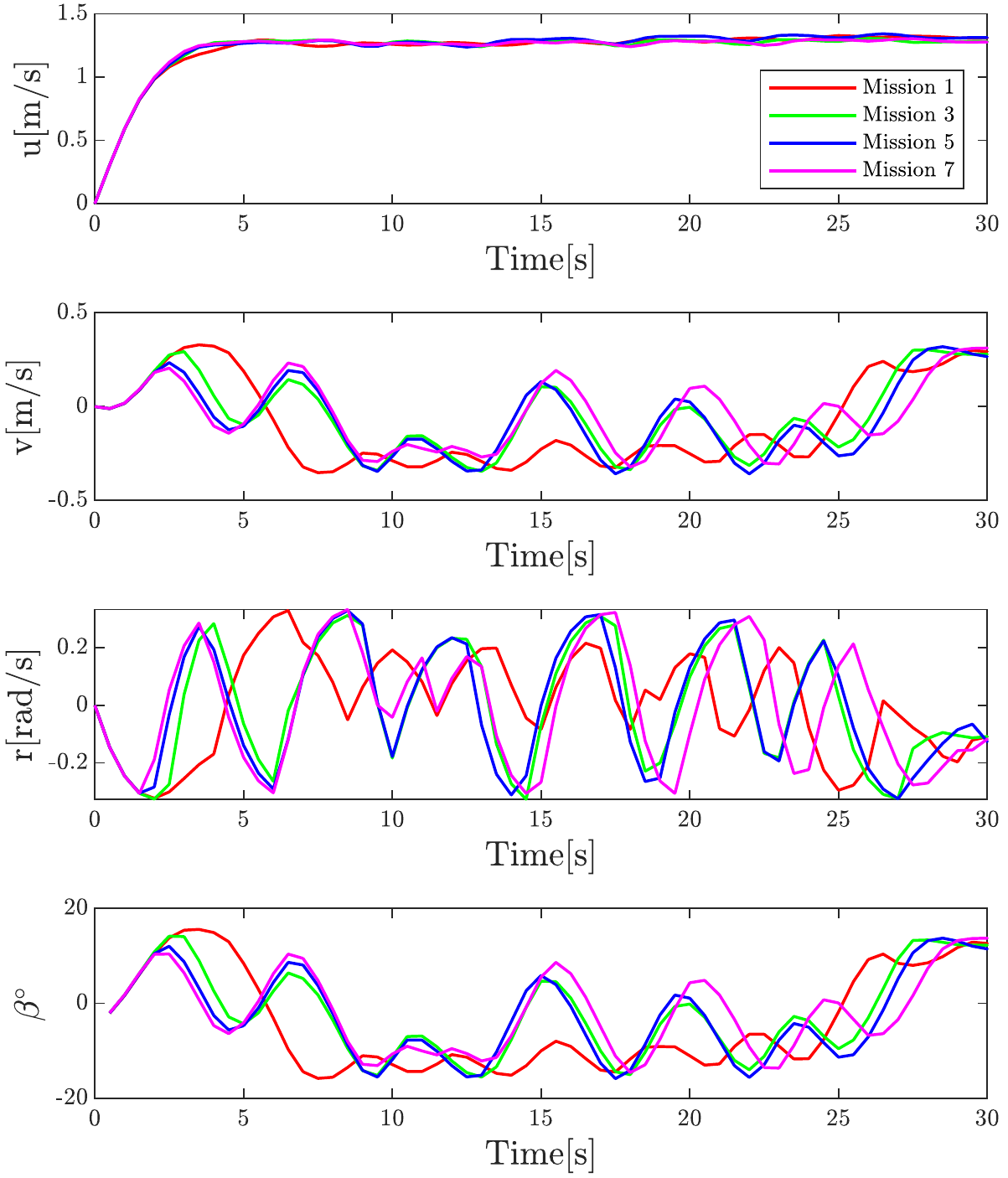}
	\caption{The surge $u$, sway $v$ and yaw $r$ velocities and sideslip angle $\beta$.}
	\label{result3}
\end{figure}

Fig. \ref{result3} shows the surge $u$, sway $v$ and yaw $r$ velocities over the missions. $\beta(t)=\arctan (\frac{v(t)}{u(t)})$ is the sideslip angle. The wind and ocean current disturbance and parametric uncertainties in ship's model are effective factors in increasing the absolute value of this angle.

\begin{figure}[ht!]
	\centering
	\def\svgwidth{0.3\textwidth}
	\includegraphics[width=0.48\textwidth]{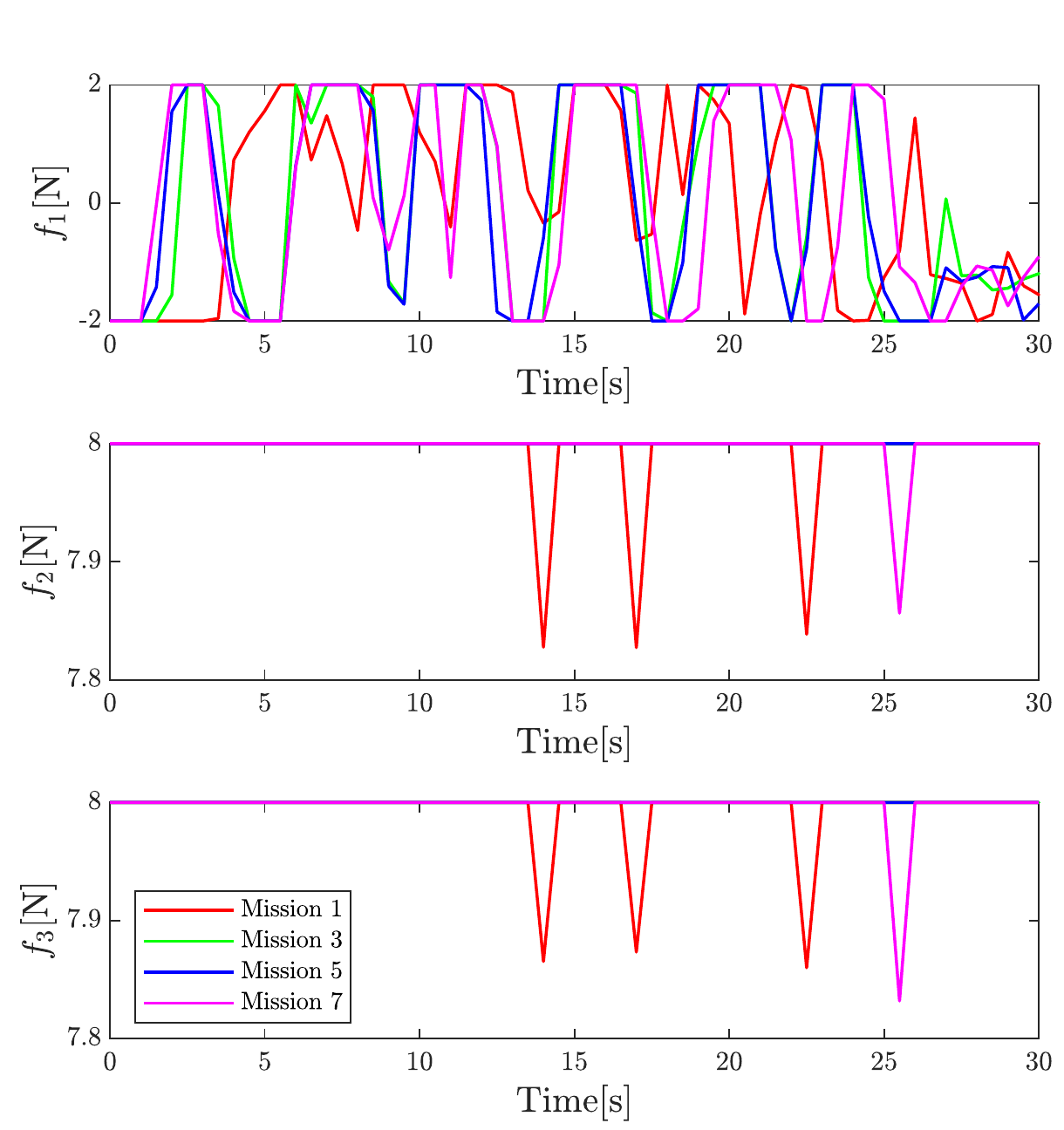}
	\caption{Actuator forces: $f_2$ and $f_3$ are the propeller, and $f_1$ is the tunnel thrusters.}
	\label{result4}
\end{figure}

The control inputs (thruster forces) are provided in Fig. \ref{result4} for the nominal system. As it is observed, the propeller thrusters $f_2$ and $f_3$ work in their upper bounds as expected to reduce the cost of route to the target point.

\begin{figure}[ht!]
	\centering
	\def\svgwidth{0.3\textwidth}
	\includegraphics[width=0.48\textwidth]{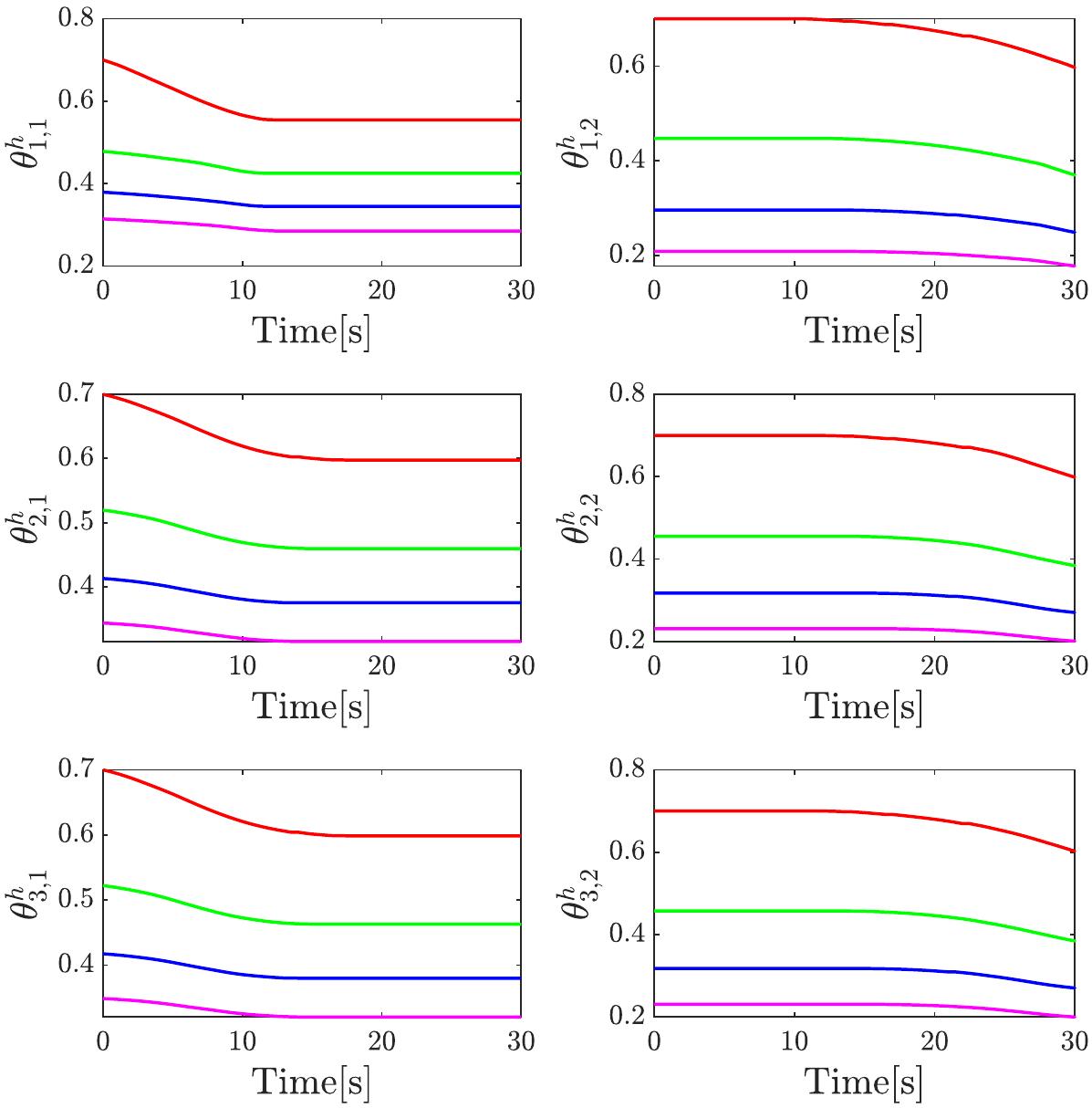}
	\caption{RL-parameters in constraint back-off $\vect \theta^h$. Red, Green, Blue and Magenta for mission 1, 3, 5, 7, respectively.}
	\label{result4.5}
\end{figure}

The back-off RL parameters $\vect \theta^h$ changes during the learning is demonstrated in Fig. \ref{result4.5}. As can be seen, in the first mission, which has a large distance from the obstacles, the parameters are reduced in order to approach the obstacles until a certain values.

\begin{figure}[ht!]
	\centering
	\def\svgwidth{0.3\textwidth}
	\includegraphics[width=0.48\textwidth]{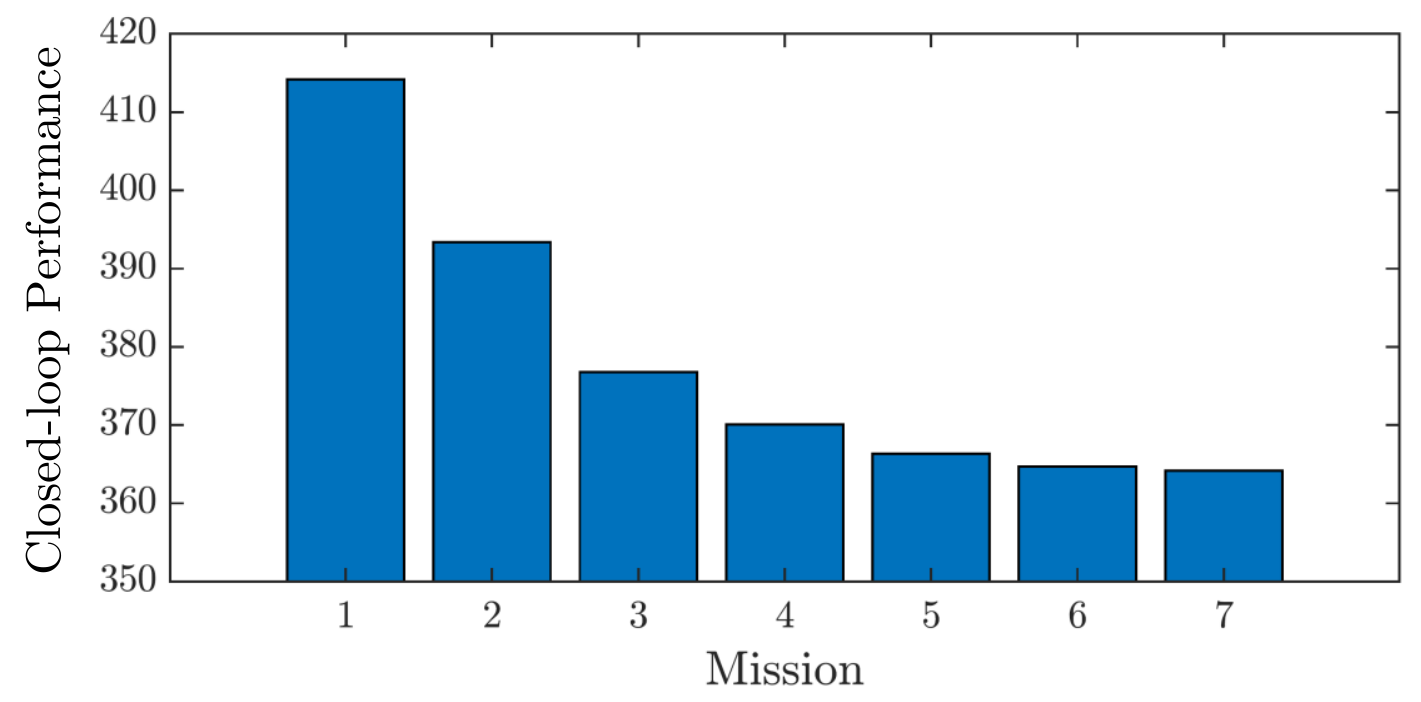}
	\caption{ Histogram of Closed-loop performance over mission}
	\label{result5}
\end{figure}

Finally, Fig. \ref{result5} illustrates the closed-loop performance of each mission. This performance is obtained by summing of baseline stage cost $L(\vect s,\vect a)$ during each episode. As can be seen, the closed-loop performance is reduced by about $12\%$ over seven missions. 
\section{Conclusion}\label{sec:conc}
 This paper proposed an RL-based RMPC technique for controling an ASV in a target tracking scenario in the presence of obstacles and stochastic wind and ocean current. A parameterized scenario-tree based MPC was used to approximate the action-value function, modelling a potential propeller thrusters failure. Additionally, constraint tightening was used in the MPC scheme to handle uncertain wind and current disturbances. The MPC tightening was adjustable by RL. A mixed energy and time cost was used as the RL's baseline cost, with the addition of a penalty when the ship trajectory was too closed to the obstacles.  We started the mission with a conservative tightening, yielding a fairly large distance from the obstacles  and let RL adjust the tightening. The simulations show how RL manages to adjust the tightening to better values. The adaptation of more parameters in the MPC scheme will be considered in the future.
 \bibliographystyle{IEEEtran}
\bibliography{shipACC}
\end{document}